\documentclass[preprint,showpacs,aps,prb,superscriptaddress,floatfix,letterpaper]{revtex4-1}

\usepackage{graphicx}
\usepackage{dcolumn}
\usepackage{bm}
\usepackage[usenames]{color} 
\usepackage{epsfig}

\newcommand{\LSMO}{La$_{2-2x}$Sr$_{1+2x}$Mn$_2$O$_7$ }
\newcommand{\LSMOn}{La$_{1.24}$Sr$_{1.76}$Mn$_2$O$_7$ }

\begin{document}

\author{C. Phatak}%
\email{cd@anl.gov} 
\affiliation{Materials Science Division, Argonne National Laboratory, 9700 S. Cass Avenue, Argonne, IL 60439, USA}%
\author{A. K. Petford-Long}%
\affiliation{Materials Science Division, Argonne National Laboratory, 9700 S. Cass Avenue, Argonne, IL 60439, USA}%
\affiliation{Dept. of Materials Science and Engineering, Northwestern University, 2220 Campus Drive, Evanston, IL 60208, USA}%
\author{H. Zheng}%
\affiliation{Materials Science Division, Argonne National Laboratory, 9700 S. Cass Avenue, Argonne, IL 60439, USA}%
\author{J. F. Mitchell}%
\affiliation{Materials Science Division, Argonne National Laboratory, 9700 S. Cass Avenue, Argonne, IL 60439, USA}%
\author{S. Rosenkranz}%
\affiliation{Materials Science Division, Argonne National Laboratory, 9700 S. Cass Avenue, Argonne, IL 60439, USA}%
\author{M. R. Norman}%
\affiliation{Materials Science Division, Argonne National Laboratory, 9700 S. Cass Avenue, Argonne, IL 60439, USA}%


%
 \title{Investigation of ferromagnetic domain behavior and phase transition at nanoscale in bilayer manganites}
 
\begin{abstract}
Understanding the underlying mechanism and phenomenology of colossal magnetoresistance in manganites has largely focused on atomic and nanoscale physics such as double exchange, phase separation, and charge order. Here we consider a more macroscopic view of manganite materials physics, reporting on the ferromagnetic domain behavior in a bilayer manganite sample with a nominal composition of \LSMO with $x=0.38$, studied using in-situ Lorentz transmission electron microscopy. The role of magnetocrystalline anisotropy on the structure of domain walls was elucidated. On cooling, magnetic domain contrast was seen to appear first at the Curie temperature within the $a-b$ plane. With further reduction in temperature, the change in area fraction of magnetic domains was used to estimate the critical exponent describing the ferromagntic phase transition. The ferromagnetic phase transition was accompanied by a distinctive nanoscale granular contrast close to the Curie temperature, which we infer to be related to the presence of ferromagnetic nanoclusters in a paramagnetic matrix, which has not yet been reported in bilayer manganites.
\end{abstract}


\maketitle

\section{Introduction}
Bilayer manganites such as \LSMO exhibit a rich phase diagram based on their doping level, which includes ferromagnetic (FM), antiferromagnetic (AF), and charge-ordered phases \cite{Argyriou1999,Ling2000}. Due to the complex crystal structure, there are several different exchange interactions within these materials that contribute to their behavior. For example the inter-bilayer exchange along the $c-$axis is weaker than the intra-bilayer exchange within the $a-b$ plane, as a result of the intrinsic two-dimensional layered structure \cite{Moritomo1996,Perring2001}. These anisotropic exchange interactions along with the competition among orbital, charge, and spin 
order, as well as lattice distortions, lead to interesting and complex magnetic and transport properties. The double exchange interaction between the Mn$^{3+}$ and Mn$^{4+}$ ions results in the material undergoing a phase transition from a paramagnetic (PM) insulator to a ferromagnetic (FM) metallic state below the Curie temperature \cite{Konishi1998}. As a result of this dramatic change in conductivity, the layered manganites exhibit a colossal magnetoresistance (CMR) effect, which has garnered much attention in the past two decades from both a fundamental as well as an applications context.
 
One of the proposed mechanisms for the CMR effect is that small FM regions, which are connected in a percolative manner \cite{Uehara1999}, form as the material is cooled through the transition temperature. Magnetic interactions and domains in these manganites and related materials have previously been studied using various techniques such as neutron scattering \cite{Osborn1998,Rosenkranz2000} and magnetic force microscopy \cite{Lu1997}, as well as Kerr microscopy \cite{Gupta1996}. The majority of the research efforts towards understanding the formation of FM domains has been done using reciprocal space and scattering methods. 
Only recently there have been some efforts towards direct real space visualization of the way in which the small FM regions form and become connected, leading to the formation of FM domains within the material below $T_c$.
 Lorentz transmission electron microscopy (LTEM) has also been used to study cubic manganites since it offers high spatial resolution and a direct visualization of the magnetic domains \cite{Asaka2002,Loudon2002,Tao2011a}. Furthermore with current advanced in-situ capabilities, LTEM offers unique possibilities to study the magnetic phase transitions as a function of temperature, while simultaneously obtaining information about structural and charge ordering using electron diffraction. Phase-reconstruction methods enable quantitative magnetic induction maps to be obtained that can provide information about the nature of the magnetic domain walls as well as physical parameters such as the exchange stiffness of the sample. 

In this work, we have explored the behavior of \LSMO with $x=0.38$, which has been reported to show a ferromagnetic transition with a Curie temperature of $T_c = 125$ K \cite{Medarde1999}. At this 
doping level, the magnetic moments in the unit cell are oriented such that the crystal has a strong easy plane ($a-b$ plane) anisotropy with $K_u \approx -2.5\times10^5$ J/m$^3$.\cite{Welp2001} The behavior of the magnetic domains and the relationship between the crystal structure and domain structure is discussed in detail together with a derivation of magnetic parameters obtained directly from the nanoscale imaging. Furthermore, we also describe the ferromagnetic phase transition and the observation of a granular nanoscale contrast that provides direct evidence of the coexistence of FM and PM phases in a bilayer manganite.

\section{Experimental methods}
Single crystals of \LSMO with $x=0.38$ i.e., \LSMOn were synthesized using the floating zone method \cite{Mitchell1997}. TEM samples were prepared from these crystals using focused ion-beam milling method as well as conventional polishing, followed by gentle milling by low energy Ar$^+$ ions to improve electron transparency. In order to fully understand the magnetic domain behavior and elucidate the role of magnetocrystalline anisotropy on the formation of domain walls, two samples were fabricated with differing geometry; (1) S1 - with the hard 
axis ($\langle001\rangle$) in the plane of the TEM sample, (2) S2 - with the hard axis ($\langle001\rangle$) perpendicular to the plane of the TEM sample. The magnetic domain behavior in the samples 
was then analyzed in the Lorentz TEM mode using a Tecnai F20 transmission electron microscope. Through-focus series of images were acquired with a nominal defocus ranging between $\Delta f = 500-1000~\mu$m. It should be noted that Lorentz microscopy is only sensitive to magnetization components that are perpendicular to the direction of the electron beam. The local magnetization was analyzed using the gradient of the phase shift of 
electrons passing through the sample. This phase shift was recovered using the {\it transport-of-intensity} equation method \cite{Volkov2002}. {\it In situ} experiments were performed using a liquid N$_2$ stage that is capable of cooling the sample to $90$ K, in order to observe the magnetic domain behavior during the magnetic phase transition from the paramagnetic state to the ferromagnetic state.

\section{Results}
\subsection{Magnetic domain walls}
\begin{figure}[h]
\centering\leavevmode
\epsffile{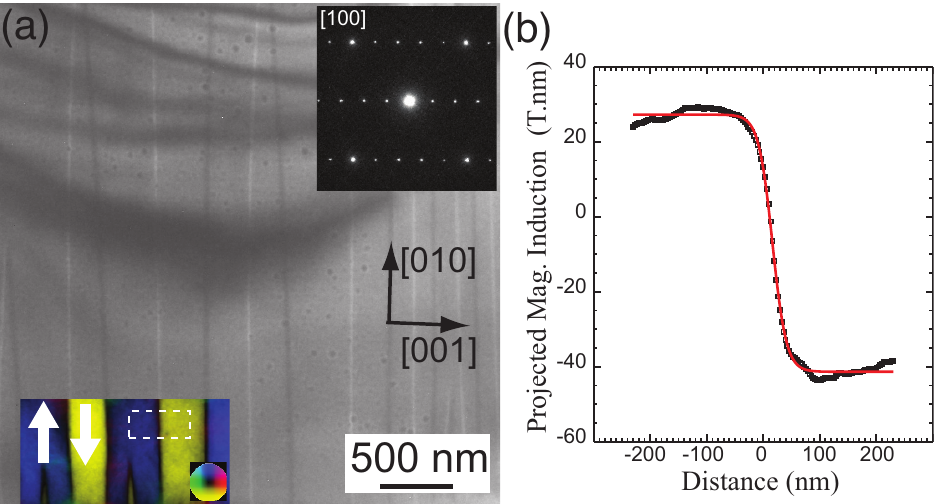}
\caption{\label{fig:domains_xc}(Color online)((a) shows the under-focused LTEM image of sample S1 at $95$ K. The top right inset shows the diffraction pattern along the $\langle100\rangle$ zone axis and the schematic shows the orientation of the crystallographic axes. The bottom left inset shows the magnetization color map overlaid on the image showing the presence of $180^\circ$ domain walls. (b) shows the plot of the projected magnetic induction (black squares) across the domain wall computed by averaging the values shown in the dashed region in (a) and a fit obtained to determine the domain wall width (red). }
\end{figure}

Figure~\ref{fig:domains_xc}(a) shows 
an under-focused Lorentz TEM image from sample S1 with the hard magnetic axis, $\langle001\rangle$, in the plane of the TEM sample. The inset (top right) shows the diffraction pattern viewed along the $\langle100\rangle$ zone axis and the orientation of the crystallographic axes in the sample plane is indicated. The sample was cooled to $95$ K, which is well below the Curie temperature.
As expected, $180^\circ$ domain walls are present, seen as bright and dark sharp lines, 
running vertically in the image. The magnetization map within this region was reconstructed from the phase shift of the electrons and is shown as a color map overlaid on the bottom left of the image. The color indicates the direction of magnetization as given by the color wheel. 
The additional curved lines seen running horizontally in the image are bend contours, which are related to strong electron diffraction effects.  
This composition of \LSMO is expected to have an easy plane anisotropy, which means that the magnetization prefers to lie in the $a-b$ plane. Due to the specific geometry of this TEM sample and its crystallographic orientation, we are observing these $a-b$ planes edge on, thereby effectively creating a strong uniaxial anisotropy in the TEM sample, with domain walls separating domains running perpendicular to the $\langle001\rangle$ direction. This also manifests itself via the formation of needle-like domains seen in the magnetization color map near the bottom of the image, which is also the edge of the sample. This type of domain is formed in order to minimize the stray field energy. The widths of the domains near the edge and inside the sample are determined by a balance between the domain wall energy and the closure (stray) field energy. The domain pattern observed here is an example of two-phase branching, which refines the domain pattern near an edge \cite{Hubert:1998kq}. This effect is observed in sample S1 because it has a strong effective uniaxial anisotropy along the $\langle010\rangle$ direction resulting in the magnetization lying along only two easy magnetization directions: $[010]$ and $[0\bar{1}0]$.
 
Since the hard axis for magnetization is in the plane of the sample, the domain walls can be expected to be of Bloch type where the magnetization rotates out-of-plane across the wall. 
The width of the domain wall can be related to physical constants such as the exchange stiffness and magnetocrystalline anisotropy using the relation: $\delta \sim \pi\sqrt{A/|K_u|}$ \cite{Hubert:1998kq}. Using the classical approximation of spin rotation across a $180^\circ$ domain wall, the distribution of the in-plane component of the magnetic induction can be approximated using the relation: 
\begin{equation}\label{eqn:bfit}
B_y = a + b\tanh\{\pi(x-c)/\delta\}, 
\end{equation}
where $a,b,c$ are constants and $\delta$ is the domain wall width. Figure~\ref{fig:domains_xc}(b) shows a plot of the in-plane component of the projected magnetic induction across the domain wall (black squares). The values were averaged over the region showed by dashed lines in Figure~\ref{fig:domains_xc}(a). A non-linear least-squares fit to the measured data was performed (shown in red) using equation \ref{eqn:bfit}, from which, the domain wall width was determined to be $77$ nm. Furthermore, using the value of $K_u = 2.5\times10^5$ J/m$^3$ from the literature \cite{Welp2001}, a value for the exchange stiffness constant for \LSMOn was determined to be $A = 1.45 \times 10^{-10}$ J/m. This demonstrates that we can determine the magnetic parameters of a material directly using nanoscale imaging. The exchange stiffness constant can be related to the exchange interactions and is dependent on the crystal structure of the material. The relationship is well established for cubic materials but not for bilayer manganites. 

\begin{figure}[ht]
\centering\leavevmode
\epsffile{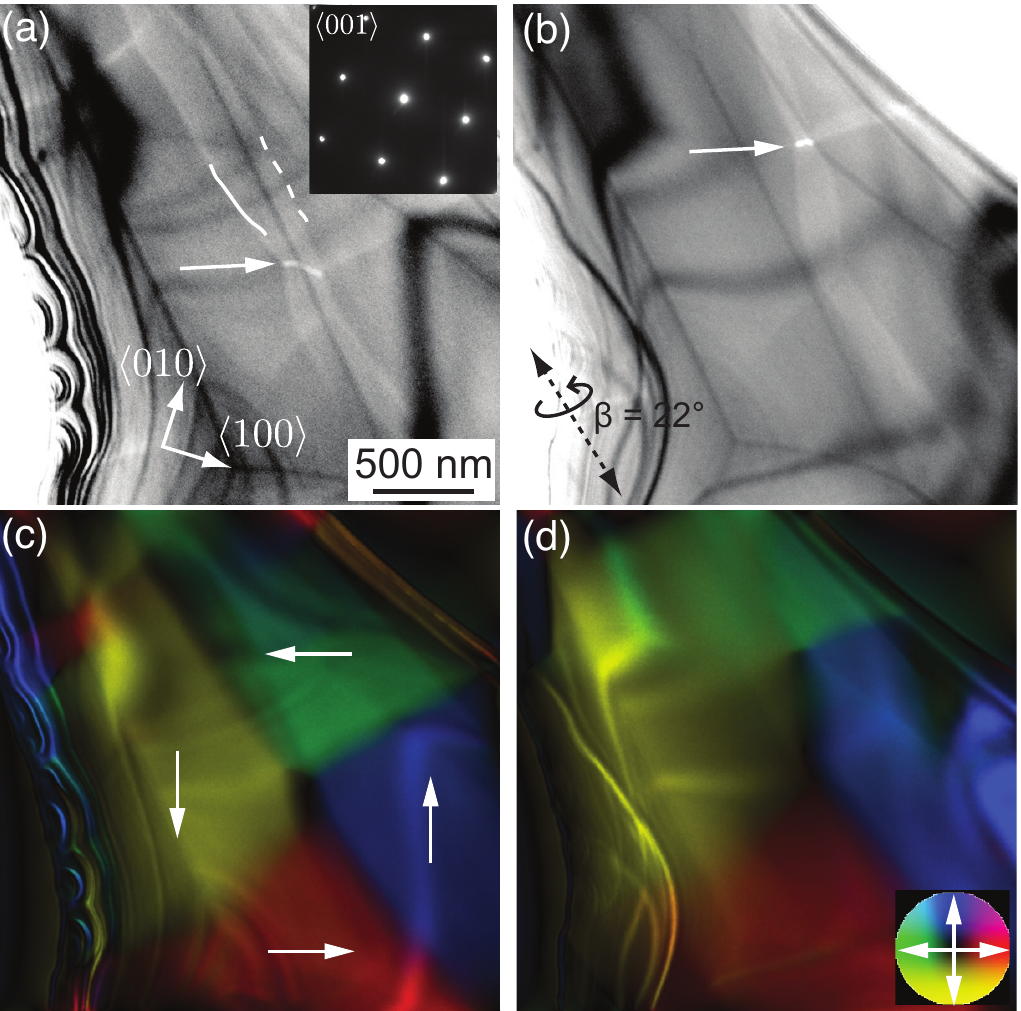}
\caption{\label{fig:domains_pl} (Color online) (a) shows the under-focused LTEM image from sample S2 at $95$ K. The inset shows the diffraction 
pattern obtained along the $\langle001\rangle$ zone axis and the orientation of the crystallographic axes is shown schematically. (b) shows the underfocus 
LTEM image of the same region after tilting the sample by $22^\circ$ about the axis shown in the figure. (c) and (d) show the reconstructed 
magnetization color map for (a) and (b) respectively. The colorwheel indicates the direction of magnetization.}
\end{figure}

The magnetic domain structure in sample S2, which has the hard axis of magnetization perpendicular to the plane of the sample, is shown in Figure~\ref{fig:domains_pl}.
Figure~\ref{fig:domains_pl}(a) shows an under-focused LTEM image from this sample. The top-right inset shows the diffraction pattern along the $
\langle001\rangle$ zone axis and the in-plane crystallographic directions are indicated. The domain walls are not seen as sharp lines as they were 
for sample S1, but now show a broad band-like contrast as highlighted by the white lines. The width of the band-like contrast varies from narrow at the edge of the sample to broad inside the sample. 
In this orientation, the easy plane ($a-b$) of magnetization is in the plane of the TEM sample, and the surface termination and sample edges lead 
to formation of a closure domain configuration to minimize the stray fields. This is clearly seen from the colored magnetization map shown in 
Figure~\ref{fig:domains_pl}(c). The magnetization direction within each region is close to a $\langle110\rangle$ type direction. It has previously 
been estimated from bulk magnetic measurements that although there is an easy plane anisotropy in \LSMOn, there is a small uniaxial 
anisotropy of about $7\times10^3$ J/m$^3$ along the 
$\langle110\rangle$ direction \cite{Welp2001}.  
It is interesting to note that at the location where two of the domain walls 
intersect, a bright white line contrast is observed (indicated by the arrow). The broadening of the domain walls can be attributed to either a large domain wall width or the presence of inclined domain walls. As we have already estimated that the domain wall width for this material is $77$ nm from the images of Sample S1, this cannot explain the broad contrast extending over a range of $300$ nm. Hence we can infer that the domain walls must be inclined with respect to the viewing direction. This was further investigated by tilting the sample to observe the effect on the domain wall contrast. 
Figure~\ref{fig:domains_pl}(b) shows an under-focused Lorentz TEM image of the same region after tilting by $22^\circ$ about the axis shown in (b). Figure~\ref{fig:domains_pl}(d) shows the corresponding colored
magnetization map. The effective broadening of the domain wall contrast has decreased along with a decrease in the length of the bright white line contrast.

\begin{figure}[h]
\centering\leavevmode
\epsffile{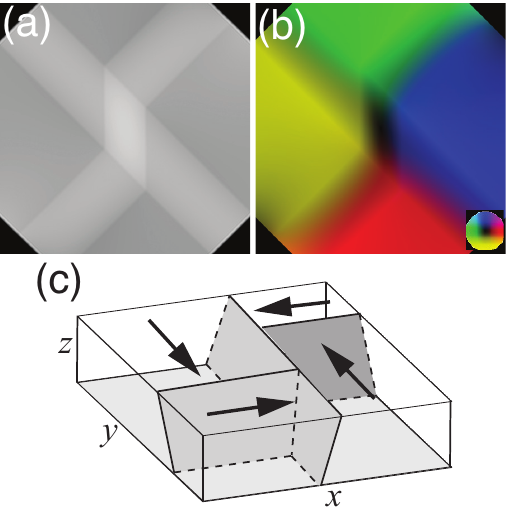}
\caption{\label{fig:simdomains_pl} (Color online)(a) shows the simulated under-focus LTEM image and (b) shows the corresponding magnetization 
color map for a model with inclined domain walls forming a closure domain configuration as shown schematically in (c).}
\end{figure}

In order to confirm the origin of the contrast, we performed image simulations as shown in Figure~\ref{fig:simdomains_pl}. Figure~\ref{fig:simdomains_pl}(a) shows a simulated underfocus image and (b) shows the corresponding colored magnetization map. The magnetic configuration with inclined domain walls (gray) used for these image simulations is shown schematically in Figure.~\ref{fig:simdomains_pl}(c). There is excellent agreement between the simulated images and the experimental ones, corroborating our view that the domain walls observed for this sample are indeed inclined with respect to the viewing direction ($\sim\langle001\rangle$). By comparison with the model, we can interpret the features indicated by the solid and dashed line in Fig.~\ref{fig:domains_pl}(a) as the intersection of the domain wall with the top surface and bottom surface of the TEM sample, respectively.

\subsection{Ferromagnetic transition}
\begin{figure}[h!]
\centering\leavevmode
\epsffile{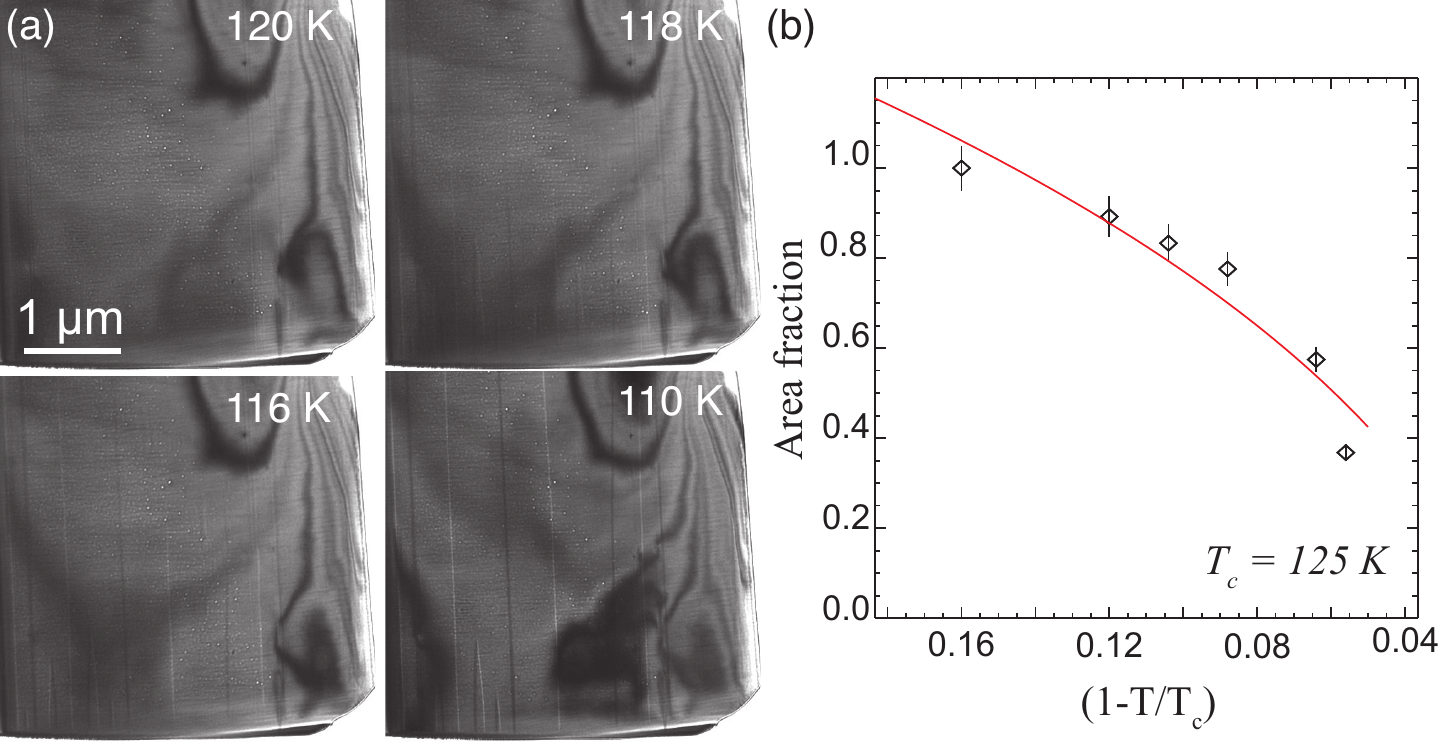}
\caption{\label{fig:insitu_xc} (a) shows a series of under-focus LTEM images acquired during the cooling of sample S1 from $120$ K to $110$ 
K. (b) shows a plot of the area fraction of magnetic domains as a function of reduced temperature (1-T/T$_c$) as calculated from the in-situ cooling image series 
(diamonds) and a power-law fit to the data (red line).}
\end{figure}

Next we explored the magnetic domain behavior as a function of temperature across the phase transition from paramagnetic to ferromagnetic state for both the sample geometries. 
Figure~\ref{fig:insitu_xc} shows the phase transition for TEM sample S1 (hard axis in the plane of the sample). As the temperature decreases from $120$ K, the 
$180^\circ$ domain walls are seen to nucleate at the edge of the sample (bottom of the images) and then grow across the TEM sample. The first appearance of magnetic domain wall contrast was observed at $T = 118$ K. This temperature is about $7$ K lower than the Curie temperature of the same sample as measured from magnetometry to be $T_c = 125$ K. This difference can attributed to the fact that at temperatures very close to $T_c$, the ferromagnetic domain signal is too weak to be detected using Lorentz TEM. Similar differences between temperature at which magnetic contrast is observed and the Curie temperature have previously been reported \cite{Loudon2006,Tao2011a}. The area fraction of the sample that was ferromagnetic was calculated as a function of temperature from this series of images. The area fraction roughly corresponds to the total magnetization of the sample under the assumption that it is uniform through the thickness of the sample. A power-law fit to the area fraction (which is representative of the magnetization) and the reduced temperature, $t = (1-T/T_c)$, using the relation $A \propto t^\beta$ yields the exponent, $\beta = 0.36$. Figure.~\ref{fig:insitu_xc}(b) shows the plot of the measured area fraction as a function of temperature (symbols) together with the power-law best fit to the data (red line). This value of $\beta$ is close to the literature reported value of $\beta=0.32$ for a three-dimensional Ising model \cite{Pelissetto2002}. Previous reports have determined the value of $\beta = 0.13$ which indicate that the phase transition below $T_c$ is still explained by the 2D Ising model \cite{Osborn1998}, however a crossover to three-dimensional scaling close to $T_c$ has been also been suggested \cite{Rosenkranz2000}.

\begin{figure}[t]
\centering\leavevmode
\epsffile{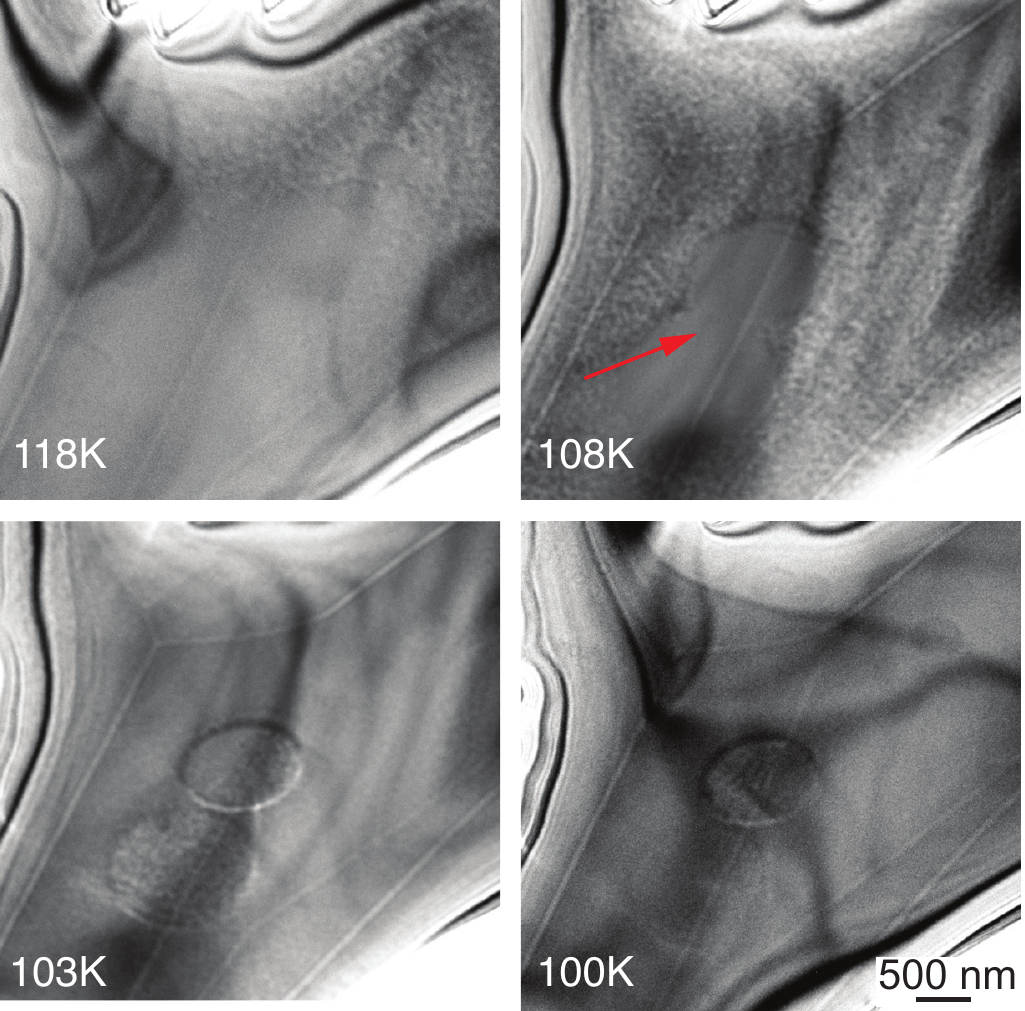}
\caption{\label{fig:insitu_pl} shows a series of under-focus LTEM images from sample S2 during in-situ cooling from $120$ K to $100$ K. The 
nanoscale granular contrast starts to appear at $118$ K and eventually disappears, leaving magnetic domain walls at $100$ K. }
\end{figure}

As for the domain wall structure, a distinctive difference was observed during the phase transition for sample S2 compared with that for S1. Figure~\ref{fig:insitu_pl} shows a series of under-focused Lorentz TEM images during cooling to below $T_c$. As the sample is cooled, there is no immediate formation of 
magnetic domain walls, but rather the formation of a nanoscale granular contrast starting from $T = 118$ K, which increases in density as the temperature decreases. The granular nanoscale contrast was only observed in the out-of-focus images and not in the in-focus image, indicating that it is magnetic in origin. Eventually these nanoscale magnetic clusters merge together to form magnetic domains separated by domain walls, leading to a decrease 
in the total number of clusters. Finally at $100$ K, most of the 
nanoscale clusters disappear leaving behind domain walls that form a closure domain configuration to minimize the stray field energy. A movie showing the {\it in situ} cooling of the sample from two different regions is 
included in the supplementary information \cite{supp-info}.
It should also be noted that at $T = 108$ K, there is a region in the center of the sample marked by red arrow that does not show any black and 
white granular contrast related to the nanoscale clusters, although it is surrounded by this contrast. 
Eventually at $T = 103$ K, the granular contrast is seen inside the region, which slowly disappears by $T = 100$ K. This suggests that there are local inhomogenieties (for example due to strain in the sample) that can result in a difference in Curie temperature. The effect of such local inhomogenieties is often missed in bulk measurements as they are averaged over the entire sample. However using LTEM, we are able to observe the coexistence of sub-micron size regions that are non-ferromagnetic in the surrounding ferromagnetic region. Similar coexistence of charge-ordered (insulating) and charge-disordered (metallic FM) domains has been previously observed in La$_{5/8-y}$Pr$_y$Ca$_{3/8}$MnO$_3$ \cite{Uehara1999}.   
Additionally as the sample was cooled, the bend contour contrast in the TEM sample was seen to change sharply over a narrow temperature range just above the Curie temperature, indicating a change in the strain state of the sample. This can be directly related with the magnetostriction of the sample as it undergoes the phase transition from the PM to FM phase. Note that the bend contour contrast stays stable over the rest of the temperature range analyzed. The abrupt change in volume and the resulting magnetostriction effect at $T_c$ has previously been reported in bilayer manganites and is associated with the insulator-to metal transition in these materials \cite{Matsukawa2002}.

\begin{figure}[ht]
\centering\leavevmode
\epsfxsize=3.3 in
\epsffile{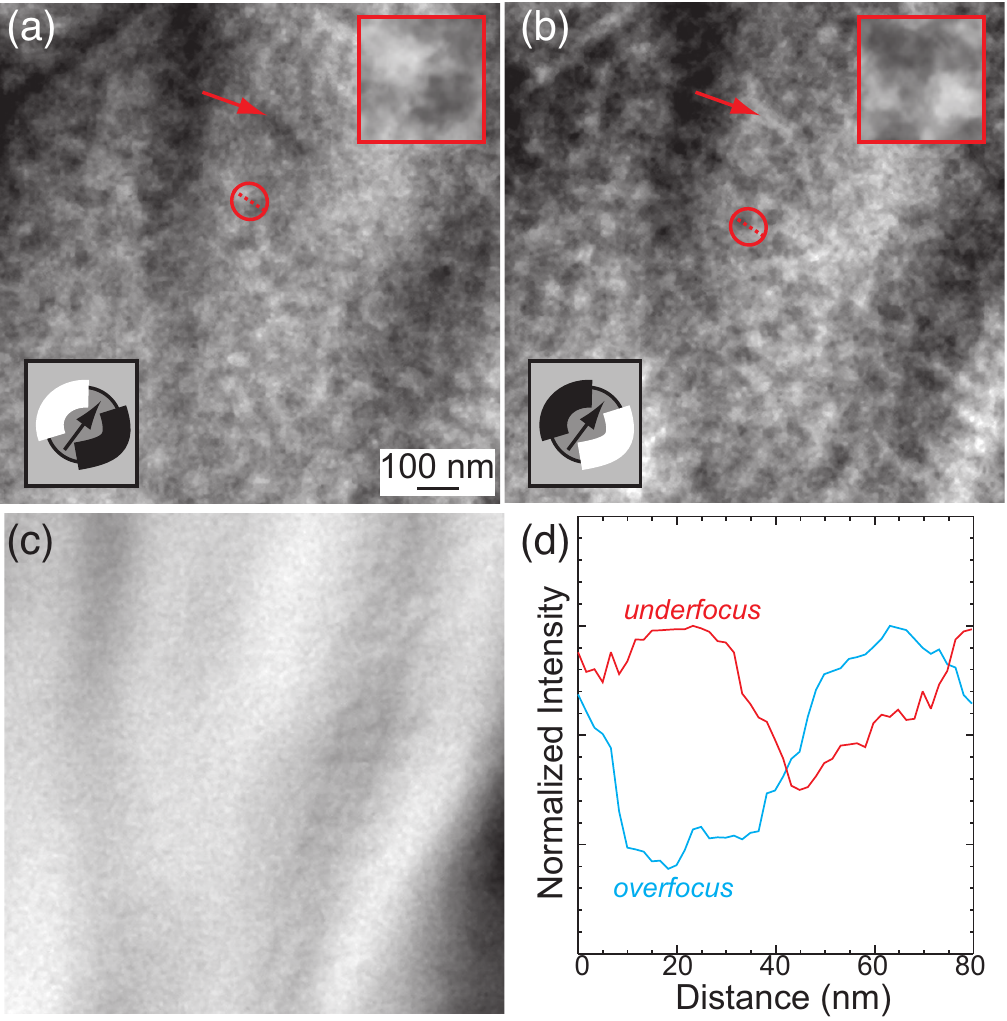}
\caption{\label{fig:nanocluster} (a)-(c) show the under-focus, over-focus and in-focus LTEM images respectively of an area showing the nanoscale 
granular contrast at $109$ K from sample S2. A magnified view of the region in red circle is shown in the top left inset showing the black and white contrast 
associated with the nanoclusters. The schematic in the bottom right inset shows the relation between the black and white contrast and the 
magnetization of the local cluster. (d) shows the plot of the normalized intensity across the dashed red line in (a) and (b).}
\end{figure}

Figure~\ref{fig:nanocluster}(a), 
(b) and (c) show the under-focus, over-focus, and in-focus Lorentz TEM images respectively of the granular contrast for $T=108$ K. The granular contrast arising from the nanoscale clusters (highlighted by the red circle)
shows a distinctive white and black intensity on either side of each cluster. The inset at the top right of Figure~\ref{fig:nanocluster}(a) and (b) shows the magnified view of the region circled in red. This black and white intensity reverses between the under-focus
and over-focus images as shown by the plot of normalized intensity in Figure~\ref{fig:nanocluster}(d), and disappears for the in-focus images. 
This type of contrast is observed for a spatial distribution of finite objects in the sample that lead to a phase shift of the electron wave passing through it, resulting in the observation of the contrast only in out-of-focus images. Thus the contrast could be related to a distribution of magnetic objects or to effects such as strain related to the phase transition. If the origin of the contrast was crystallographic, i.e. strain, then changes in the bend contour contrast would also be expected. However this was only observed prior to the appearance of the granular contrast as mentioned earlier. We therefore infer that the origin is magnetic and is evidence for the formation of a random distribution of ferromagnetic clusters in a non-magnetic matrix. Since the spins of individual atoms within these ferromagnetic clusters are aligned, each cluster can be described as a nanoscale single domain magnetic object. The expected contrast in the out-of-focus images that is associated with such a single domain magnetic object is schematically shown in the bottom-inset of Figure~\ref{fig:nanocluster}(a) and (b). Further evidence for this interpretation comes from the fact that the nanoclusters eventually merge to form domains. An example of a wall segment that has formed is indicated
by the red arrow in Figure~\ref{fig:nanocluster}(a) and (b).

A similar granular contrast of nanoclusters has previously been observed, although only in cubic manganites such as Nd$_{0.5}$Sr$_{0.5}$MnO$_3$ \cite{Asaka2002} and La$_{0.55}$Ca$_{0.45}$MnO$_3$ \cite{Tao2011a}. In both cases, the granular contrast was associated with the presence of ferromagnetic nanoclusters. However, in the case of Nd$_{0.5}$Sr$_{0.5}$MnO$_3$, the granular constrast was observed only during the phase transition from the AF phase to FM phase. In the case of La$_{0.55}$Ca$_{0.45}$MnO$_3$, the ferromagnetic nanoclusters with an ordered superstructure were seen to form within a matrix that was already ferromagnetic with sub-micron size magnetic domains. 
Here we have observed the formation of these nanoclusters in bilayer manganites during both cooling through the Curie temperature as well as heating through it, without the presence of a charge-ordered phase or any other form of superstructure. The lack of any structural or long-range charge ordering was confirmed using electron diffraction during the heating and cooling. From the plot of the intensity (Figure~\ref{fig:nanocluster}(d)), the size of these nanoclusters can be measured as 
roughly $40$ nm (peak to peak distance). However, it must be noted that the high defocus value used in these images results in additional magnification. Therefore the true size of these nanoclusters is expected to be smaller than $40$ nm. This size is still significantly larger than the lattice spacing in the $a-b$ plane of $\sim 0.4$ nm or the inter-bilayer distance of $\sim 2$ nm. This suggests that we are only able to image the clusters once they reach a size that their net magnetic moment is detectable using Lorentz TEM.

\section{Summary}
In summary, we have studied the magnetic domain wall structure in \LSMOn in the ferromagnetic regime and its relation to the crystallography of the sample. Using the freedom to prepare the TEM sample along different crystallographic orientations, we investigated the detailed structure of the domain walls and were able to conclude that fabrication of the TEM sample does not significantly alter the domain wall behavior as compared to the bulk. When the hard axis of magnetization was in the plane of the TEM sample, $180^\circ$ Bloch walls are observed. By measuring the domain wall width from the nanoscale imaging, we determined the exchange stiffness of the material. In the sample with the hard axis of magnetization perpendicular to the sample plane, we observed broad band-like contrast for the domain walls. By comparing the experimental images with simulated ones, we were able to conclude that the domain walls are inclined which results in the broadening of the contrast. By analyzing the in-situ growth of magnetic domains as a function of temperature during cooling, we were able to determine the nature of the ferromagnetic transition by fitting a power law to the magnetization versus temperature data and estimating the critical exponent $\beta$ to be $0.36$. We infer that this corresponds to a crossover to three-dimensional scaling close to $T_c$. 
We were also able to visualize the formation of nanoclusters during the phase transition close to $T=T_c$ which showed a direct evidence of co-existence of magnetic and non-magnetic phases in bilayer manganites. Additionally, we also observed that there are local sub-micron scale regions which become ferromagnetic at slightly different temperatures as compared to their surroundings. Both the formation of nanoclusters and sub-micron scale regions suggest that this phase transition is percolative in nature. 
Further detailed image analysis of the nanoclusters to determine their relative size, and density as a function of temperature could yield more insights into the details of the phase transition.

{\it Note added in proof.}  Recent work by Bryant {\it et. al.}\cite{Bryant2015}, reported on imaging the magnetic domain walls as a function of temperature in La$_{1.2}$Sr$_{1.8}$Mn$_2$O$_7$ $(x=0.40)$ using low temperature MFM. They measured $T_c$ close to $118$ K, however, they observed that the magnetic domain walls disappear at about $20$ K below $T_c$. This observation could be related to suppression of the magnetization at the surface which has been previously reported \cite{Freeland2007}. However, they are only able to observe surface effects and do not report on the formation of magnetic domain walls or the formation of nanoclusters close to $T_c$.

\begin{acknowledgments}
This work was supported by the U.S. Department of Energy, Office of Science, Materials Sciences and Engineering Division. Use of Center for 
Nanoscale Materials
 was supported by the U.S. Department of Energy, Office of Science, Office of Basic Energy Sciences, under contract
 no. DE-AC02-06CH11357.
\end{acknowledgments}

\bibliography{lsmo_lit.bib}

\newpage \section*{Figures}

\newpage

\newpage

\newpage

\newpage

\newpage


\end{document}